\documentclass[reprint,showpacs,preprintnumbers, API,prb,]{revtex4-1}
\usepackage{graphicx}
\usepackage{dcolumn}
\usepackage{bm}

\preprint{API/123-QED}

\begin{document}

\title{Intrinsic Spin-Orbit Interaction in Graphene}
\author{B. S. Kandemir}
\date{\today}

\begin{abstract}
In graphene, we report the first theoretical demonstration of how the
intrinsic spin orbit interaction can be deduced from the theory and how it
can be controlled by tuning a uniform magnetic field, and/or by changing the
strength of a long range Coulomb like impurity (adatom), as well as gap
parameter. In the impurity context, we find that intrinsic spin-orbit
interaction energy may be enhanced by increasing the strength of magnetic
field and/or by decreasing the band gap mass term. Additionally, it may be
strongly enhanced by increasing the impurity strength. Furthermore, from the
proposal of Kane and Mele [Phys. Rev. Lett. \textbf{95}, 226801 (2005)], it
was discussed that the pristine graphene has a quantized spin Hall effect
regime where the Rashba type spin orbit interaction term is smaller than
that of intrinsic one. Our analysis suggest the nonexistence of such a
regime in the ground state of flat graphene.
\end{abstract}

\maketitle

\homepage{kandemir@science.ankara.edu.tr}

\affiliation{Department of Physics, Faculty of Sciences, Ankara University, 06100\\
Tando\u{g}an, Ankara, Turkey}

The spin-orbit effect is purely kinematic effect, and arises due to the
interaction of the orbital and spin internal degrees of freedom. Recently,
there has been growing interest in the study of this effect in graphene, due
the fact that the progress in manipulating graphene\cite{NOVOSELOV2004}, and
graphene based nanostructures, particularly graphene quantum dots \cite%
{CHEN2007,RYCERZ2007,BADARSON2009} has opened new perspectives in usage of
spin qubits \cite{LOSS1998} for quantum computing and quantum information
purposes \cite{RECHER2010}. Graphene based nanostructures seem to be proper
candidate for realizing these kind of processes through controlling one or
two basic spin interactions such as intrinsic or Rashba type spin-orbit
interactions (SOIs) \cite%
{KANE2005,HARNENDO2006,MIN2006,KUEMMETH2008,RASHBA2009,
HARNENDO2009,NETO2009,GMITRA2009,RAKYTA2010,SANCHO2011,ABANIN2011,BLIOKH2011,LIU2012}%
. Since the former is supposed to be small, investigations are focused on
the latter one which assigns a finite mass for the graphene carriers due to
the interaction with substrate, and thus leading to a finite gap. In fact,
within the Spin Quantum Hall Effect (SQHE) context, investigations were
initiated by Kane and Mele's work\cite{KANE2005}. They proposed that the
ground state of graphene exhibits a SQHE. They showed that SQHE regime
exists provided that Rashba type SOI term, $\Delta _{\mathrm{SO}}^{\mathrm{R}%
}\sim 0.05\ \mathrm{meV}$, smaller than that found for intrinsic SOI, $%
\Delta _{\mathrm{SO}}^{\mathrm{I}}\sim 0.103\ \mathrm{meV}$. Later, on the
basis of microscopic considerations, $\Delta _{\mathrm{SO}}^{\mathrm{I}%
}(\Delta _{\mathrm{SO}}^{\mathrm{R}})$ is estimated to be much smaller
(larger) than that of found by Kane and Mele: For instance, Hernando \textit{%
et a}l\cite{HARNENDO2009} found $\Delta _{\mathrm{SO}}^{\mathrm{I}}\sim
0.86\ \mu \mathrm{eV}$, Min \textit{et a}l\cite{MIN2006} found $\Delta _{%
\mathrm{SO}}^{\mathrm{I}}\sim 0.57\ \mu \mathrm{eV}$, $\Delta _{\mathrm{SO}%
}^{\mathrm{R}}\sim 0.011\ \mathrm{meV}$. Nevertheless, it is concluded that
SQHE regime can occur only below a temperature $\Delta _{\mathrm{SO}}^{%
\mathrm{I}}\sim 0.01\ K$ ($10^{-3}\ \mathrm{meV}$). In the impurity context,
very recently Castro and Neto\cite{NETO2009} showed that impurity induced
distortion in flat graphene lead to a significant enhancement of SOI about $%
\Delta _{\mathrm{SO}}^{\mathrm{IMP}}\sim 1-7\ \mathrm{meV}$. While our
analysis confirms this estimate within the pure impurity coverage, but it
reveals that, in the absence of impurities, i.e., in the case of flat
graphene, there is no experimentally accessible temperature where SQHE
regime can occur.

Since intrinsic SOI is a natural property of Dirac's equation, i.e., it
inherently involves the spin degrees of freedom, its Graphene's analog
contains pseudospin or just spin-orbit interaction automatically. By
graphene's spin analog, we refer two equivalent points in the Brillouin
zone, i.e., $\mathrm{K}$ and $\mathrm{K}^{^{\prime }}$ valleys as pseudospin
or just spin\cite{Mecklenburg2011}. In this letter, we first propose a
simple model to address the question of how intrinsic SOI energy in graphene
can be obtained, and then can be controlled by the strength of applied
magnetic field, and/or impurity coverage as well as gap parameter.

In graphene, near the Dirac point, the electronic states in the presence of
electromagnetic potential $\ A_{\mu }=\left( A_{0},\mathbf{A}\right) $ are
described by the effective low-energy Dirac equation, $\mathcal{H}\Psi
=E\Psi $ where
\begin{equation}
\mathcal{H}=v_{F}\boldsymbol{\alpha }\cdot \left( \mathbf{p}-\frac{e}{c}%
\mathbf{A}\right) +\beta Mv_{F}^{2}-eA_{0}  \label{11}
\end{equation}%
is the Dirac Hamiltonian. Here, $\mathbf{A}$ and $A_{0}$ are the vectoral
and scalar parts of the four vector potential $A_{\mu }$, respectively, and
they will be chosen as $\mathbf{A=B}(-y,x,0)/2$ in the Coulomb gauge, and $%
Ze/\epsilon r$ for charged Coulomb impurity, respectively. $\boldsymbol{%
\alpha }_{i}$ and $\boldsymbol{\beta }_{i}$ are Dirac matrices, and $%
v_{F}\simeq 10^{6}{m}/{s}$ is the Fermi velocity. In the absence of external
fields, the energy eigenvalues of $\mathcal{H}$ yield $\overline{E}=\pm
\sqrt{\overline{M}_{0}^{2}+\overline{k}_{0}^{2}}$ which determines linear
dispersion $\overline{k}_{0}\left( \mathbf{-}\overline{k}_{0}\right) $ for
the conduction (valence) in the absence of a gap term $\overline{M}_{0}={a}%
M_{0}={2}\overline{{\Delta }}/3${. We have used abbreviations }$%
M_{0}=Mv_{F}/\hbar ={2}\overline{{\Delta }}/3a $ together with definitions ${%
\hbar v_{F}=3}aJ_{0}/2$, where $a$ and $J_{0}$ are the carbon-carbon
distance ($1.42{\mathring{A}}$) and transfer integral ($\sim 2.7eV$) between
them, respectively. Thus, $\overline{E}={E}{a}/{\hbar v_{F}}$ and $\overline{%
k}={k}{a}$ are dimensionless energy and wave vector. Throughout the letter,
we restrict ourselves in a single valley ( $\mathrm{K} $ ) and in a single
band (conduction).

By successively applying Fouldy-Wouthuysen unitary transformation on
wave functions and operators of Dirac equation $\mathcal{H}\Psi =E\Psi $, one
passes the two component equation at any desired order of electromagnetic
interaction strength, and thus the small and large components of $\Psi $ are
completely decoupled\cite{MESSIAH1961}. Hence, by just restricting ourselves
to the positive energy solutions, together with the corresponding upper
components of $\Psi $, expansion of $\mathcal{H}$ in power of the strength
of electromagnetic interaction, as it should be, yields the well-known two
component Pauli equation, $\mathcal{H}^{\mathrm{FW}}\Phi =(Mv_{F}^{2}+%
\mathcal{H}_{\mathrm{n.r.}})\Phi $ where
\begin{equation}
\mathcal{H}_{\mathrm{n.r.}}=\frac{1}{2{m}}\left( \mathbf{p}-\frac{e}{c}%
\mathbf{A}\right) ^{2}-\frac{{e}\hbar }{2{m}{c}}\boldsymbol{\sigma }\cdot
\boldsymbol{B}-{e}{A_{0}}  \label{22}
\end{equation}%
whose eigenvalue equation $\mathcal{H}_{\mathrm{n.r.}}\Phi =E_{\mathrm{n.r.}%
}\Phi $ yields well-known nonrelativistic results, but in the two components
formalism. Continuing the expansion to second order, one can obtain three
more terms that we represent them as $\Delta $, thus, $\mathcal{H}^{\mathrm{%
FW}}\Phi =(Mv_{F}^{2}+\mathcal{H}_{\mathrm{n.r.}}+\Delta )\Phi $. These are
relativistic dependence of kinetic energy, SOI energy, and the Darwin terms,
respectively. The last term is different from zero at where there are
charges creating the field \cite{BERESTETSKII1982}. Furthermore, it is
ascribed to Zitterbewegung\cite{BJORKEN1964}, and may be dominated at higher
magnetic fields. Thus, just by taking the difference of energy spectra of $%
\mathcal{H}$ and $\mathcal{H}^{\mathrm{FW}}$ given by Eq.~(\ref{11}) and
Eq.~(\ref{22}), respectively, we are left with SOI energy. In other words,
it may be defined as $\Delta E_{\mathrm{SO}}=-Mv_{F}^{2}+E-E_{\mathrm{n.r.}}$%
, where $E$ is the relativistic energy spectrum. In dimensionless form, it
can be rewritten as $\Delta \overline{E}_{\mathrm{SO}}=-\overline{M}_{0}+%
\overline{E}-\overline{E}_{\mathrm{n.r.}}$.

We first discuss the influence of magnetic field on the graphene electronic
energy spectrum in the presence of a single Coulomb impurity. In this case,
the Dirac Hamiltonian is by
\begin{equation}
\mathcal{H}=\mathcal{H}_{0}-\hbar v_{F}\frac{e}{c}\boldsymbol{\sigma }\cdot
\mathbf{A},  \label{33}
\end{equation}%
\quad where\quad
\[
\mathcal{H}_{0}=\mathbf{-}i\hbar v_{F}\left( \boldsymbol{\sigma }\cdot
\boldsymbol{\nabla }\right) -\left( {Z}{e}^{2}/\epsilon {r}\right)
\]
is the exactly solvable unperturbed part\cite{NOVIKOV2007} whose energy
eigenvalues are easily found to be as $\overline{\epsilon }_{n}=\overline{M}%
_{0}\left[ 1+\overline{Z}^{2}/\left( n+\gamma \right) ^{2}\right] ^{-1/2}$
together with the corresponding eigenfunctions in terms of Laguerre
polynomials

\begin{equation}
\Psi _{nj}(r)=\frac{1}{\sqrt{r}}\left(
\begin{array}{rl}
& F_{nj}\left( r\right) e^{i\left( j-1\right) } \\
i & G_{nj}\left( r\right) e^{i\left( j\right) }%
\end{array}%
\right)  \label{1}
\end{equation}%
where%
\begin{eqnarray*}
\left.
\begin{array}{c}
F_{nj}\left( r\right) \\
G_{nj}\left( r\right)%
\end{array}%
\right\} &=&\left( -1\right) ^{n}N_{nj}\left( \overline{Z},M_{0}\right)
\sqrt{M_{0}\pm \epsilon _{n}} \\
&&e^{-\lambda r}\left( 2\lambda r\right) ^{\gamma -1/2}\left[ L_{n}^{2\gamma
}\left( 2\lambda r\right) \pm C_{21}L_{n-1}^{2\gamma }\left( 2\lambda
r\right) \right]
\end{eqnarray*}%
with%
\[
N_{nj}\left( \overline{Z},M_{0}\right) =\left\{ \frac{\Gamma \left(
n+1\right) \lambda ^{3}\left[ m_{j}+\left( \overline{Z}M_{0}/\lambda \right) %
\right] }{\Gamma \left( n+2\gamma +1\right) \overline{Z}M_{0}^{2}}\right\}
^{1/2}
\]%
is the normalization constant, and $\ \ C_{21}\ =-\left( n+2\gamma \right) /%
\left[ m_{j}+\left( \overline{Z}M_{0}/\lambda \right) \right] $, $\ \lambda =%
\sqrt{M_{0}^{2}-\epsilon _{n}^{2}}$, $\gamma =\sqrt{j^{2}-\overline{Z}^{2}}$%
, and $L_{n}^{2\gamma }$ are Laguerre polynomials. Here, $\overline{Z}={Z}{e}%
^{2}/\epsilon \hbar v_{F}$ is the dimensionless coupling constant and $%
j=m_{j}+1/2$ is the eigenvalue of the conserved total angular momentum $%
J_{z}=L_{z}+S_{z}$. The values of the quantum number $n$ are $n=0,1,2,\ldots
$ if $m_{j}\geq 0$, and $n=1,2,\ldots $ if $m_{j}<0$. It can be easily check
that $\overline{\epsilon }_{0}=\overline{M}_{0}\left[ 1+\left( 2\overline{Z}%
\right) ^{2}\right] ^{1/2}$ becomes zero at $\overline{Z}_{cr}=1/2$. In the
framework of perturbation theory, by using Eq.~(\ref{1}), one obtains
first-order shift in energy eigenvalues of Eq.~(\ref{33}) as $\overline{E}=%
\overline{\epsilon }_{n}+\Delta \overline{\epsilon }_{n}$, where
\begin{equation}
\Delta \overline{\epsilon }_{n}=\frac{1}{4\overline{Z}\left( \overline{M}%
_{0}^{2}\right) }\left\{ \overline{B}\sqrt{\overline{M}_{0}^{2}-\overline{%
\epsilon }_{n}^{2}}\left( 2n+\gamma \right) j+\frac{\overline{Z}\overline{M}%
_{0}}{\sqrt{\overline{M}_{0}^{2}-\overline{\epsilon }_{n}^{2}}}\right\} ,
\label{2}
\end{equation}%
is the Zeeman term due to the magnetic field. In Eq.~(\ref{2}), $\overline{B}%
=B/B_{0}$ is the dimensionless magnetic field with $B_{0}=\hbar {c}/{e}{a}%
^{2}$. The result given in Eq.~(\ref{2}) is valid for $\overline{B}\ll
\overline{Z}^{2}\overline{M}_{0}^{2}/2$. In the same units, the energy
eigenvalues $\overline{E}_{n}$ together with its two-dimensional
non-relativistic equivalent which can easily be obtained\cite{HO2000} as
\[
\overline{E}_{\mathrm{n.r.}}=-\left( \frac{\overline{M}_{0}\overline{Z}^{2}}{%
2}\right) \frac{1}{\left[ n+\left\vert m\right\vert +\left( 1/2\right) %
\right] ^{2}}+\frac{\overline{B}\ \overline{M}_{0}}{2}\left( m+s\right)
\]%
can be used to calculate SOI energy as functions of strengths of magnetic
field, gap $\overline{{\Delta }}$ parameter as well as of impurity strength $%
\overline{Z}$.

First, in the absence of magnetic field, to show impurity strength
dependency of the SOI energy alone in the graphene, we plotted the lowest
lying states in fixed gap values as a function of $\overline{Z}$ in FIG.~\ref%
{FIG1}(a). For comparison, in the inset, we have also included a plot
showing the gap parameter dependency of SOI energy for the same energy
levels but at fixed impurity strengths. It is clear from the figure that, in
comparison with SOI energies for higher orbital, SOI energy for the
ground-state is higher, and all they are getting more pronounced with
increase in $\overline{Z}$ and/or $\overline{{\Delta }}$. The former follows
from the fact that, higher states signify larger radius so that the
corresponding impurity binding energy, and hence the SOI energy becomes
smaller. The latter can easily be approved by analyzing the coupling
constant and mass dependent form of SOI energy for two-dimensional
relativistic Hydrogen atom\cite{STRANGE1998}. As a result, not only \ our
these findings support the enhancement of SOI by impurity covarage\cite%
{NETO2009} , but they also show how it can be controlled by means of the
relevant parameters such as impurity strength and gap parameter.

\begin{figure}[tp]
\includegraphics [  height=10.27cm,width=7.cm]{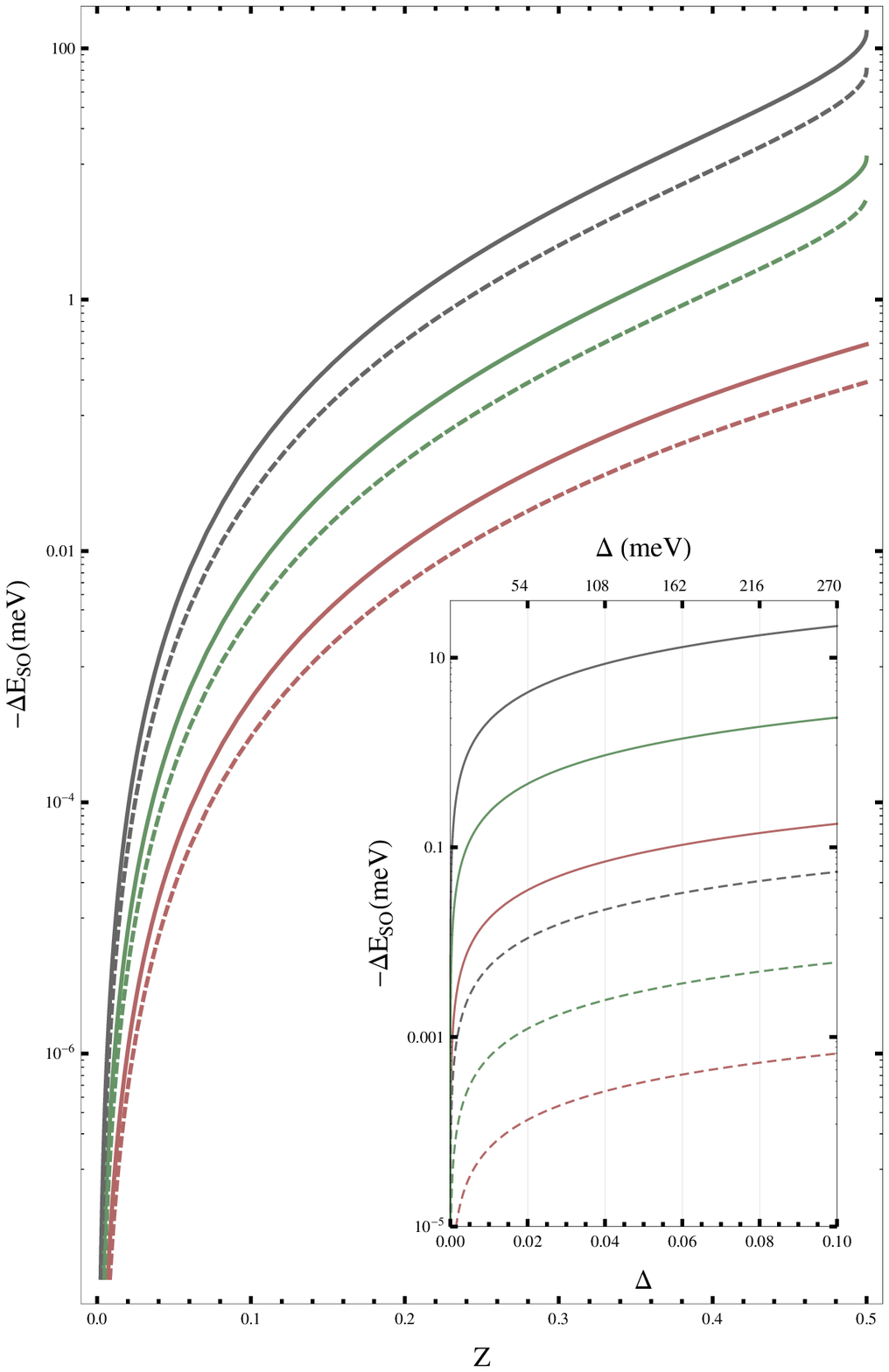} (a) %
\includegraphics [  height=10.27cm,width=7.cm]{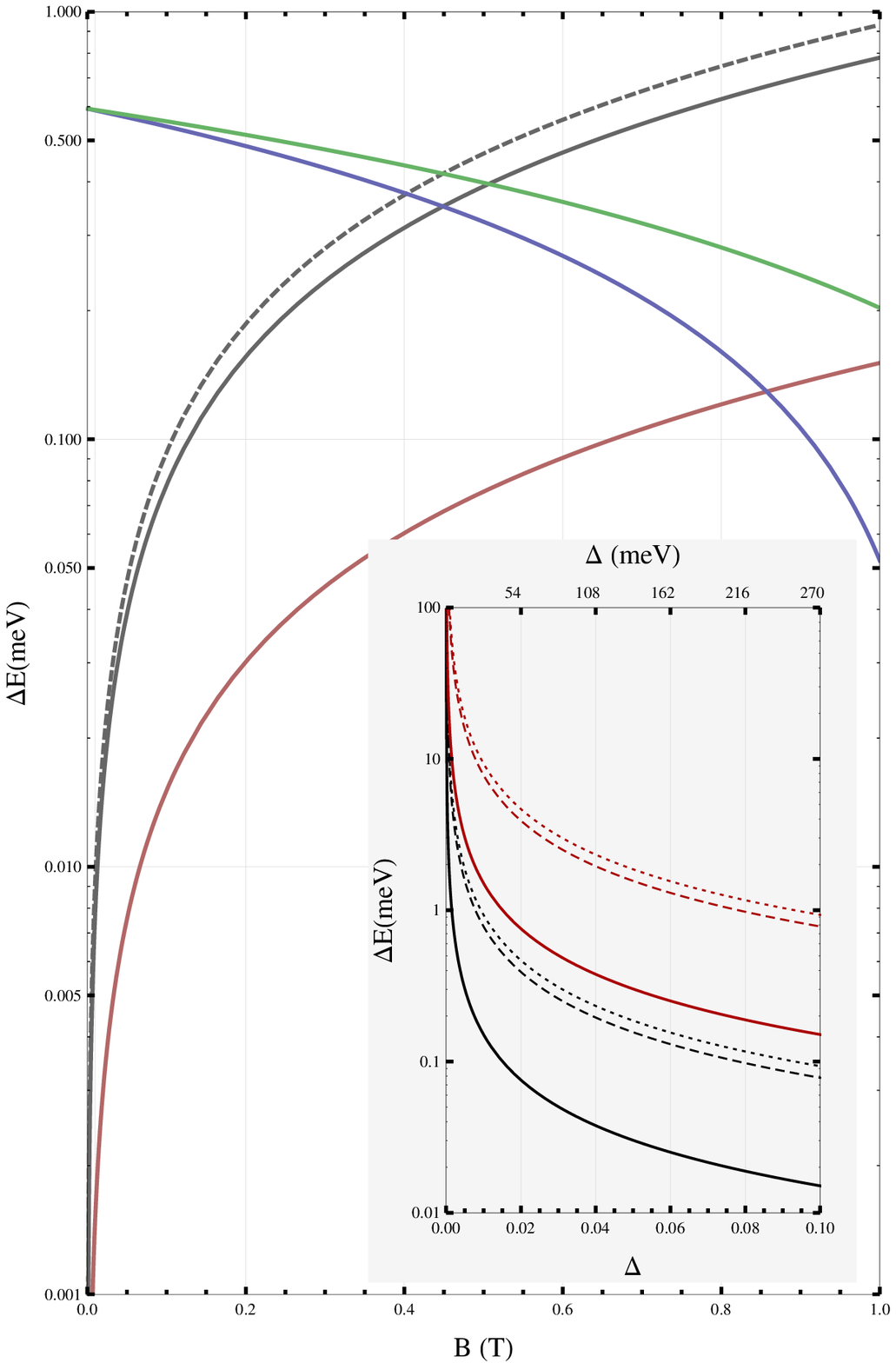} (b)
\caption{(Color online) (a) SOI energies as function of $\overline{Z}$ at
gap parameter $\overline{{\Delta }}=0.1$ (straight lines) and $\overline{{%
\Delta }}=0.01$ (dashed lines) for the three lowest-lying energy levels,
namely, from top to bottom, $[n,j]\equiv [ 0,1/2],[1,-1/2]\equiv [1,+1/2]$
and $[0,-3/2]\equiv $ $[0,3/2]$, respectively. Inset shows its dependence on
$\overline{{\Delta }}$ at $\overline{Z}$ $=0.1$ (dashed lines) and $%
\overline{Z}$ $=0.4$ (straight lines), respectively, for the same levels.
(b) Relativistic (bold black line) and non-relativistic (dashed black line)
Zeeman splitting energies in $[1,-1/2]$ , $[1,+1/2]$ levels, together with
their difference (red straight line) as function of magnetic field at fixed
value $\overline{Z}=0.3$. For this value of $\overline{Z}$ , we also plot
individual SOI for each level, i.e., $[1,-1/2]$ (green), $[1,+1/2]$ (Blue)
as function of magnetic field. Inset: relativistic (dashed lines) and
nonrelativistic (dotted lines) Zeeman energies together with their
difference (bold straight lines) as function of gap parameter for two
different magnetic field values, i.e., $B=0.1 \,\text{T}$ (black) and $1\,%
\text{T}$ (Red), respectively.}
\label{FIG1}
\end{figure}

In the presence of uniform magnetic field, as is seen from Eq.~(\ref{2}),
the orbital degeneracies are lifted. In particular, it splits the first
excited doubly degenerate $j$ levels, and thus leads to Zeeman splitting. By
increasing the magnetic field, as expected, the Zeeman splitting spacing
broadens gradually with $B$, (FIG.~\ref{FIG1}(b)). To show the combined
effects of magnetic field on SOI together with the effect of SOI on Zeeman
splitting, we plot the Zeeman splitting for the first excited state in the
relativistic spectrum as function of magnetic field in FIG.~\ref{FIG1}(b).
We also include a corresponding curve for the non-relativistic energy
spectrum as a reference. Hence, by just taking the difference between these
two, i.e., between the relativistic and the non-relativistic energies Zeeman
energies, we obtain the contribution of SOI energy to the Zeeman splitting
energy. Its dependence on $B$, magnetic field dependence of SOI for $j=1/2$
and $j=-1/2$ levels are individually displayed in FIG.~\ref{FIG1}(b) for
comparison. We can see that its influence becomes significant for decreasing
$B$, since the Zeeman splitting linearly depend on $B$ and $j$. Therefore,
the difference between these two curves is simply the SOI contribution to
the Zeeman energy. To learn the response of SOI to the applied magnetic
field, we have also plotted the individual SOI energies for each level.
Obviously, we claim that, when the associated Zeeman splitting is resolved
in an experiment, the related SOI contribution, and therefore SOI energies
for each level can be resolved. As an example, at $B=1\mathrm{T}$, our
calculation describes Zeeman energy up to $0.70\,\mathrm{meV}$ which
contains $0.15\,\mathrm{meV}$ due to SOI. One can also justify this value by
just taking the difference between the SOI energies of the related levels
which are $0.20\,\mathrm{meV}$ and $0.05\,\mathrm{meV}$ , respectively. As
for the inset of the figure, we compare $\overline{{\Delta }}$ dependency of
the SOI energies. Eventually, we give a measure of how amount of Zeeman
energy is influenced by SOI. The above values may be enhanced by just
increasing $\overline{{Z}}$ and/or $\overline{{\Delta }}$ decreasing .

A similar procedure developed above for binding in the hydrogenic sense can
also be extended to the binding in oscillatory sense, in the absence of any
impurity. In pristine graphene, massless Dirac Fermions having linear
dispersion obey Dirac-Weyl equation, and they have no nonrelativistic
analog. But, we may overcome this difficulty by just using the spectrum of
Dirac equation, and then take its limit when $\overline{{\Delta }}%
\rightarrow 0$. To do this, it is enough to recall the well-known graphene
Landau levels (LLs), i.e.,%
\[
\overline{E}=\sqrt{\overline{M}_{0}^{2}+2n/\overline{\ell }_{B}^{2}}
\]%
with $n=\nu +\left( \left\vert m\right\vert +m+s+1\right) /2$. \ Here. $%
\overline{\ell }_{B}=1/\sqrt{\overline{B}}$ is the magnetic confinement
length. It can easily be shown that its non-relativistic counterpart has
energy spectrum with
\[
\ \overline{E}_{\mathrm{n.r.}}=\frac{1}{\overline{M}_{0}\overline{\ell }%
_{B}^{2}}\left( \nu +\frac{\left\vert m\right\vert +m+1}{2}\right) +\frac{%
g\ast s}{\overline{M}_{0}\overline{\ell }_{B}^{2}}
\]
\begin{figure}[tp]
\includegraphics [  height=10.27cm,width=7.cm]{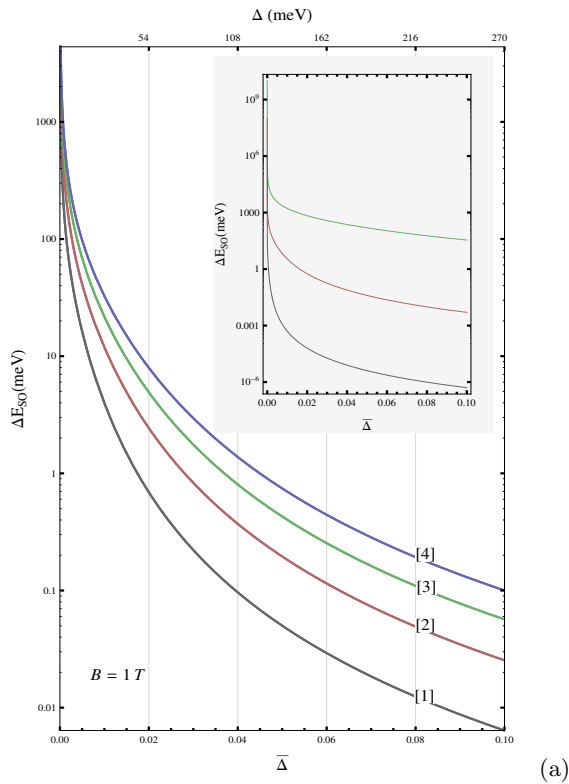} (a)
\caption{(Color online) For the four lowest eigenstates of the graphene, SOI
energies as function of gap parameter \ at $B=1\ \mathrm{T}$. The upper
scale shows the gap parameter in units of $\mathrm{meV}$. Inset: \ SOI for
free particle spectrum, i.e., $\Delta \overline{E}_{\mathrm{SO}}=-\overline{M%
}_{0}+\protect\sqrt{\overline{M}_{0}^{2}+\overline{k}_{0}^{2}}-\left(
\overline{k}_{0}^{2}/2\overline{M}_{0}\right) $ as function of gap
parameter. From top to bottom, red, green and black straight lines
corresponds $\overline{k}_{0}=0.1$, $0.01$, and $0.001$, respectively. }
\label{FIG2}
\end{figure}
wherein the first term is the well-known non-relativistic LLs, the second
one is due to Zeeman effect. Here, we will set the Land\'{e} factor to $2$.
It is easy to check analytically that SOI energy is zero for $n=0$ ($\nu =0$,%
$m=0$, $s=-1$) level even without taking $\overline{{\Delta }}\rightarrow 0$
limit (massless limit). It is also easy to see that higher orbital are
sensitive to decrease in $\overline{{\Delta }}$, i.e., $\overline{{\Delta }}%
\rightarrow 0$ limit (gapless limit). As is done in FIG.~\ref{FIG2}, it is
also instructive to compare this picture with those obtained from the free
particle spectrum, i.e., $\Delta \overline{E}_{\mathrm{SO}}=-\overline{M}%
_{0}+\sqrt{\overline{M}_{0}^{2}+\overline{k}_{0}^{2}}-\left( \overline{k}%
_{0}^{2}/2\overline{M}_{0}\right) $ which is also zero at $\overline{k}%
_{0}\rightarrow 0$ limit. Again, even in the presence of mass term, it gives
zero. Whereas it becomes more pronounced as $\overline{{\Delta }}\rightarrow
0$ for finite $\overline{k}_{0}$. The appearance of the lack of SOI in the
electron-hole degeneracy point, i.e., in both $n=0$ and $\overline{k}_{0}=0$
cases, shows that this point has purely non-relativistic character.
Furthermore, this proves that the ground-state of the flat graphene does not
exhibit SQHE regime. Additionally, beyond this point we see that SOI effects
are much more pronounced in $\overline{{\Delta }}\rightarrow 0$ limit than
those found for the impurity coverage in the same limit. These findings are
exactly compatible with the analytical predictions of the relativistic
quantum mechanics on the general form for SOI energy in general. All these
show that our predictions work well for different physical regimes.

We have three main conclusions. First, we demonstrate theoretically that,
without sophisticated many-body calculations, intrinsic SOI energy in the
electron-hole degeneracy point is zero in flat graphene. Thus, we conclude
that, for flat graphene, it is not possible to observe SQHE. Second, within
the impurity coverage, we investigate the influence of SOI on graphene
energy levels. Finally, in the same context, we investigate the Zeeman
splitting in doped graphene as well as the influence of SOI on this
splitting . We demonstrate that it is possible to enhance intrinsic SOI by
increasing the strength of applied magnetic field, and how it can be
resolved from the Zeeman effect. Therefore, magnetic field not only can be
used to tune the Zeeman effect, but also allow us to estimate the magnitude
of the SOI energy in dopped graphene. As for the Zitterbewegung, its effect
should be taken account for higher levels and higher magnetic field values.
Its signature is the oscillatory behavior of the velocity and for its direct
observation stronger magnetic fields would be required \cite{Romera2009}. It
can also be resolved in the same procedure described here within this
context, i.e., but for higher levels where SOI is negligible, and for higher
magnetic fields. We hope that this study will not just reveal the intrinsic
SOI in graphene, but also it will illuminate the Berry phase in graphene,
since very recently it is shown that, by constructing the intrinsic SOI
operator from the non-Abelian Berry connection, intrinsic SOI originates
from Berry phase terms \cite{Bliokh2011}.

\begin{acknowledgments}
I thank Professor T. Altanhan for valuable discussions.
\end{acknowledgments}

\end{document}